\documentclass[preprint,aps,showpacs]{revtex4}
\usepackage[ulem=normalem]{changes}
\usepackage{graphicx}

\usepackage{subfigure}
\usepackage{dcolumn}
\usepackage[T1]{fontenc}
\usepackage{bm}
\usepackage{amsmath}
\usepackage{amsfonts}
\usepackage{float}
\usepackage{latexsym}
\usepackage{amssymb}
\usepackage{epsfig}
\usepackage{color}
\usepackage{graphicx}
\usepackage{parskip}
\usepackage{tabularx}

\def \beq{\begin{equation}}

\def \eeq{\end{equation}}
\def \bea{\begin{eqnarray}}
\def \eea{\end{eqnarray}}

\def \bs{\boldsymbol}

\begin{document}
\title{Identifying insulating states of ultra cold atoms with cavity transmission spectrum 
}
\author{Jasleen Lugani}
\affiliation{Department of Physics, IIT Delhi, New Delhi-110016, India}
\author{K. Thyagarajan}
\affiliation{Department of Physics, IIT Delhi, New Delhi-110016, India}
\author{Sankalpa Ghosh}\email{sankalpa@physics.iitd.ac.in}
\affiliation{Department of Physics, IIT Delhi, New Delhi-110016, India}
\begin{abstract}
In this paper, we consider the transmission characteristics of an optical cavity loaded with ultra cold atoms in a one dimensional optical lattice
at absolute zero temperature. In particular, we consider the situation when the many body quantum state of the ultra cold atoms is 
an insulating state with fixed number of atoms at each site, which can be either density 
wave (DW) or  Mott insulator (MI) phase, each showing different type of discrete lattice translational symmetry. We provide a 
general framework of understanding the transmission spectrum from a single and two cavities/modes
loaded with such insulating phases. Further, we also discuss how such a transmission spectrum changes when these insulating phases 
make a cross over to the superfluid (SF) phase with the changing depth of the optical lattice potential. 
\end{abstract}
\maketitle
\section{Basic introduction}
\label{Intro}
Various quantum phases of ultra cold atoms cooled near the absolute zero temperature provide a unique opportunity 
to study a large number of theoretically envisaged exciting quantum many body states in a highly controlled situation
\cite{dalibard}. The phenomenal development in this field was made possible due to the  development in laser cooling and trapping techniques in the earlier decades \cite{Cohen} that exploit the laser-atom interaction in a novel way. Such laser-atom interaction,  besides creating a confining potential, also provides a highly controlled external potential that 
modifies the property of the quantum many body states in a profound way. 

The most common method for the experimental detection of such quantum many body states is to allow the atomic cloud to expand after switching off the laser-atom interaction and then to take the time of flight (TOF) image of the same through absorption spectroscopy \cite{BEC, Greiner}.  For example, a superfluid (SF) 
state having phase coherence but uncertainty in atom number distribution, will show a clear diffraction pattern in its TOF image. In contrast, a Mott insulator (MI) phase 
with no phase coherence, but having equal number of atoms at each site will show no such maxima and minima in the resulting TOF image.

However, in the last few years there has been a tremendous progress in other detection techniques. One of the most noteworthy of these alternative techniques involves cavity quantum electrodynamics with ultra cold atoms \cite{Ritsch}. Here a collection of ultra cold atoms is loaded in a high-finesse Fabry Perot cavity and the atoms are allowed to interact with the selected cavity mode(s). The presence of cavity enhances the atom-photon interaction and the coupled equations of the atomic operator and the cavity mode operator need to be solved simultaneously, in order to obtain the atom and field dynamics. 
Due to this atom-photon coupling, the cavity transmission 
spectrum carries information of the many body quantum mechanical state of the cold atoms loaded inside the cavity. For example, 
the cavity transmission spectrum corresponding to MI state  and an  SF state will be very distinct and hence 
these phases can be differentiated  without making a direct measurement on the atomic system
 \cite{mekhov1}. This observation makes cavity quantum electrodynamics, a complementary 
technique to the well known time of flight (TOF) imaging technique that was earlier used 
to probe such MI to SF transition \cite{Jaksch, Greiner} and led to a significant experimental development in the field of cavity  quantum electrodynamics /cavity optomechanics 
with ultra cold atoms \cite{Bren1, Colom1, Zimmerman1, Botter}.

Almost all earlier theoretical studies on ultra cold atoms in optical lattices loaded inside a cavity, mostly 
consider  MI and SF phases and the phase transition them~\cite{Ritsch,mekhov1,larson1,larson2}, which assumes short range interaction among such cold atoms. 
However, in recent years, a number of experimental systems have been explored that have additionally long range interactions \cite{dipole1,polar1,Grimm, Lev}. 
Such systems in an optical lattice can be modeled as extended Bose Hubbard model (EBHM).  In EBHM, apart from the typical onsite interaction between the cold atoms, there is also  nearest neighbor interaction (NNI)
at the minimal level and in general, can have higher order long range interactions such as  next nearest neighbor  etc.\cite{ebhm1,ebhm2}. 
The inclusion of NNI and higher order interactions in the usual Bose Hubbard model leads to the formation of  new exotic insulating phases such as 
various density wave (DW) phases, and also supersolid (SS) phase which is an intermediate phase between such DW phase and the SF phase \cite{supersolid}. Unlike 
the extensive study on the transmission spectra from a cavity loaded with MI or SF phases of ultra cold atoms, the nature of the cavity transmission spectrum 
in the presence of other exotic phases such as DW or SS phases has not yet taken place been explored. 

Motivated by this, in this work, we go beyond 
the MI and SF phase \cite{mekhov1}, and show 
how cavity transmission spectrum can distinguish between a general class of insulating phases and their transition to superfluid phase. 
We emphasize that while in TOF imaging technique, all insulating phases show the same type of diffraction pattern due to the  
absence of phase coherence, our suggested generalization of the measurement technique of cavity transmission can probe the respective 
lattice translational symmetry of such insulating phases. We also introduce a simple but instructive ansatz in the form a superposition of such general 
insulating phase and a superfluid phase and study the behavior of the cavity transmission as a function of the variational parameter. 
This further helps us to understand the transition from the respective insulating phase to the superfluid phase using cavity transmission spectra. 

The organization of the paper is as follows: In Sec.~\ref{theory}, we briefly derive the Hamiltonian used to describe such a system and illustrate the method used for finding out the exact wavefunction of the many body atomic system. In Sec.~\ref{scm}, we present the results for the transmission spectra using the methodology described in Sec.~\ref{theory} for a single cavity mode. We present various translational and angle resolved measurements for different kinds of insulating states and also compare DW to SF phase transition with the MI to SF phase transition. In Sec.~\ref{tcm} we present the results for the normal mode splitting for different phases, considering two cavity modes. Finally in Sec.~\ref{conclu}, we conclude our results with possible experimental implications. 

\section{Theoretical Framework}
\label{theory}
\begin{figure}[h!]
\includegraphics[scale=0.8]{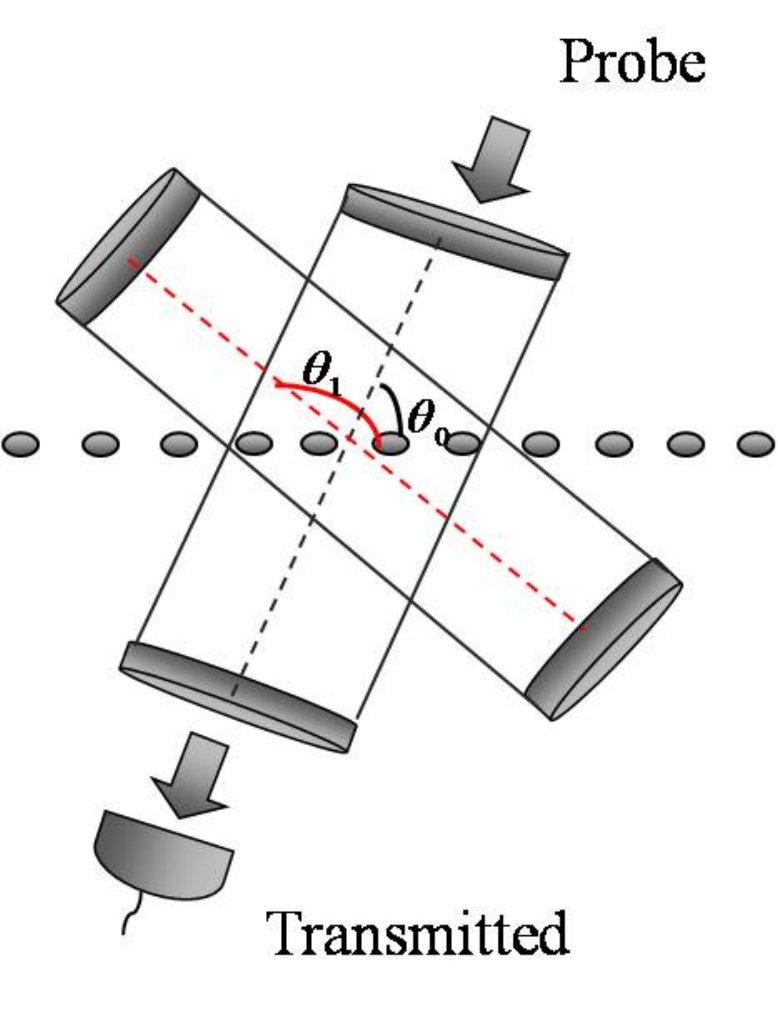}
\caption{(Color online) Schematic of the model, a system of $N$ cold atoms at $M$ sites (represented as gray shaded circles), illuminated by a cavity, pumped by an external probe. Transmission spectrum of the two cavities can be studied by placing a detector at the other end of the cavities. In the present case, only one of the cavities (inclined at an angle $\theta_0$, with respect to the lattice axis) is pumped through the external probe, and the photons from this cavity are scattered to the second cavity (inclined at an angle $\theta_1$ with respect to the lattice axis).}
\label{scheme}
\end{figure}
The physical system under consideration is schematically shown in Fig. \ref{scheme}. It consists of 
$N$ two level bosonic atoms at $M$ lattice sites in a prototype 
one dimensional optical lattice formed by external lasers. 
In such an optical lattice the atoms scatter light in different directions. To enhance the atom-light interaction along a particular direction, 
such atoms in optical lattice are placed inside a cavity and a particular 
cavity mode is excited with the help of an external pumping laser. The interaction of the quantum many body state of 
the ultra cold atoms with this cavity mode is considered by analyzing the transmitted  photons from the 
cavity. Such a transmission spectrum  carries signature of the quantum many body state of the cold 
atoms loaded in the cavity. A number of recent experiments \cite{Bren1, Colom1, Zimmerman1, Bren2, Gupta} 
have realized such experimental systems and paved the way of doing cavity quantum electrodynamics with ultra cold atoms. 

The theoretical framework for studying the photon transmission spectrum from a cavity loaded with ultra cold atoms in optical lattice 
was introduced in a seminal paper by the Innsbruck group \cite{mekhov1} which was subsequently expanded and elaborated in  \cite{mekhov2,mekhov3,mekhov4,mekhov5,mekhov6,mekhov7,mekhov8}. The developments were recently reviewed in \cite{mekhovreview}.  In this framework, in the tight binding approximation,   
if we restrict the interaction between the neighboring atoms upto the nearest neighbors, the extended Bose-Hubbard model \cite{Damski} that includes the cavity mode-atom interaction can be obtained as~\cite{mekhov4}
\bea  
\hat{H}&=&\sum_{l} \hbar \omega_{l} \hat{a}_{l}^{\dag} \hat{a}_{l} \hspace{-1pt} - i\hbar \sum_{l} (\eta_{l}^{*}(t) \hat{a}_{l} - \eta_{l}(t) \hat{a}_{l}^{\dag}) + J_{0}^{cl} N + J^{cl} \sum_{j=1}^{M} \hat{b}_{j}^{\dag}\hat{ b}_{j+1}  +
 \hbar g^{2} \sum_{l,m} \frac{\hat{a}_{l}^{\dag} \hat{a}_{m}}{\Delta_{ma}} \Bigg( \sum_{j=1}^{K} J_{j,j}^{lm} \hat{n}_{j} \Bigg) \nonumber \\ &+& \hbar g^{2} \sum_{l,m} \frac{\hat{a}_{l}^{\dag} \hat{a}_{m}}{\Delta_{ma}} \Bigg( \sum_{<j,k>}^{K} J_{j,k}^{lm}\hat{ b}_{j}^{\dag}\hat{ b}_{k} 
\Bigg) + \frac{U}{2} \sum_{j=1}^{M} \hat{n}_{j} (\hat{n}_{j} -1) +V \sum_{j=1}^{M} \hat{n}_{j}\hat{n}_{j+1}
\label{ebhm} \eea
The above Hamiltonian has been derived under rotating wave approximation in which all the fast
oscillating terms in the Hamiltonian are neglected, and adiabatic elimination of the excited state of atoms~\cite{mekhov1} (assuming far off resonant interaction between atoms and the cavity photons). The first term in Eq. (\ref{ebhm}) represents the free field Hamiltonian of the cavity modes (the index $l$ runs over the number of cavity modes, $l$=0 corresponds to single cavity mode and $l$=1 corresponds to the second cavity mode), the second term represents the interaction between the external pump and the cavity mode. $\hat{a}_{l}$ is the annihilation operator of the cavity mode with the frequency $\omega_{l}$, wave vector $ {\bf{k}}_{l}$, and mode function $u_{l}({\bf{r}})$. $\eta_{l}(t) = \eta e^{-i\omega_{p}t}$ is the time dependent amplitude of the external pump laser of frequency $\omega_{p}$ that populates the cavity mode.

In the third term of Eq. (\ref{ebhm}), $J_{0}^{cl}$ corresponds to the matrix element of the atomic Hamiltonian in the site localized Wannier basis, $ w({\bf{r}} - {\bf{r}}_{j}) $ and represents the onsite interaction energy amplitude, given as
\begin{equation}
J_{0}^{cl} = \int d{\bf{r}}w({\bf{r}} - {\bf{r}}_{j})  [ -\frac{\hbar^{2} \nabla^{2}}{2m_{a}} + V_{cl}({\bf{r}})]
 w({\bf{r}} - {\bf{r}}_{j}) 
\end{equation} 
where, $-\frac{\hbar^{2} \nabla^{2}}{2m_{a}}$ represents the kinetic energy of the atoms. Similarly, $J^{cl}$ represents the nearest neighbor hopping amplitude in the fourth term, given as
\begin{equation}
J^{cl} = \int d{\bf{r}}w({\bf{r}} - {\bf{r}}_{j})  [ -\frac{\hbar^{2} \nabla^{2}}{2m_{a}} + V_{cl}({\bf{r}})]
 w({\bf{r}} - {\bf{r}}_{j+1}) 
\end{equation}
$\hat{b}_{j}$ is the bosonic 
annihilation operator at site $j$ and $ \hat{n}_{j} = \hat{b}_{j}^{\dag} \hat{b}_{j} $ represents the atom number operator at lattice site $j$. The term $V_{cl}(\bs{r})$ includes the classical optical lattice potential that arises due to the dipolar interaction of the ultra cold atoms with the oscillating electric field of the lasers that form the optical lattice. This also includes any other external trapping for such cold atoms. 
The fifth and the sixth terms are the contributions from the atom-cavity mode interaction to the onsite and the hopping amplitude, respectively, where $J_{j,k}^{lm}$ is the matrix element given as 
\begin{equation}
J_{j,k}^{lm} = \int d{\bf{r}}w({\bf{r}} - {\bf{r}}_{j})  u_{l}^{*}({\bf{r}}) u_{m}({\bf{r}}) w({\bf{r}} - {\bf{r}}_{k}), 
\end{equation}
$\Delta_{la}$  = $\omega_{l}$ - $\omega_{a}$ denotes the cavity atom detunings, where $\omega_{a}$ is the frequency corresponding to the energy level separation of the two-level atoms and $ g $ is the atom-light coupling constant.
$U$ and $V$ are the interaction matrix elements in the Wannier basis, originated from the two body 
interaction potential \cite{Damski}. This Bose Hubbard model is different from the one originally proposed by Jaksch \textit{et al}
\cite{Jaksch, Greiner} since it includes the interaction between the atomic state and the cavity mode, apart from the classical optical lattice potential included in $V_{cl}(\bs{r})$ . $M$ is the total number of sites and out of which $K$ sites are illuminated ($K\leq M$).

 For a sufficiently deep optical lattice potential included in  $V_{cl}({\bf{r}})$, the overlap between Wannier functions can be neglected and they can be approximated as delta functions. In this limit,  $J^{cl} = 0$ and $J_{j,k}^{lm} = 0$ for $j\neq k$.
We first consider the case where a single mode $l=0$ is excited inside the cavity and the transmitted photons are also collected in the same mode. 
If one assumes that the tunneling time from one optical lattice site to another is much longer than the interaction time 
between the cavity photons and ultra cold atoms, the photon induced self organization of the  quantum many body state
\cite{Gopal1, Gopal2, Vidal} can be neglected. Under this approximation, 
Heisenberg equation of motion for $\hat{a}_{0}$  gives us
\begin{equation}
\dot{\hat{a}}_{0} = -i\omega_{0}\hat{a}_{0} - i \delta_{0}\hat{D}_{00} \hat{a}_{0} - \kappa \hat{a}_{0} + \eta_{0}(t)
\label{pn}
\end{equation} 
where, $\hat{D}_{00} = \sum_{j=1}^{K} {u_{0}}^{*}({\bf{r}}_{j})u_{0}({\bf{r}}_{j})\hat{n}_{j}$, $\delta_{0} = g^{2}/ \Delta_{0a} $ and $ \kappa $ is the cavity relaxation rate. The transmission spectra for the cavity photons can be studied using the expectation value of the photon number operator which can be calculated from the above equation for the steady state situation as
\begin{equation}
\hat{a}_{0}^{\dag}\hat{a_{0}}=\frac{|\eta|^2}{(\Delta_p-\delta_0 \hat{D}_{00})^2+\kappa ^2}
\label{a0}
\end{equation}
where $\Delta_p$ is the probe-cavity detuning, the term $\delta_0 \hat{D}_{00}$ signifies the shift in the cavity resonance due to the presence of the cold atoms and reveals the nature of the many body quantum state of the ultra cold atoms 
inside the cavity.

The above treatment can be straightforwardly generalized when the mode $0$ is excited inside the cavity and the transmitted photons are collected into another mode $1$, with the corresponding annihilation operator $\hat{a}_{1}$. 
The Heisenberg equation for the mode operator $\hat{a}_{1}$ can now be found as 
\begin{equation}
\dot{\hat{a}}_{1} = -i\omega_{1}\hat{a}_{1} - i \delta_{1}\hat{D}_{11} \hat{a}_{1}- i \delta_{0}\hat{D}_{10} \hat{a}_{0} - \kappa \hat{a}_{1}
\label{2m}
\end{equation}
If for simplification, we assume that the two cavity modes $0, 1$ are of same frequency, resulting in $\omega_{1}$=$\omega_{0}$, $\delta_{1}=\delta_{0}$. $\hat{D}_{11}$=$\sum_{j=1}^{K} {u_{1}}^{*}({\bf{r}}_{j})u_{1}({\bf{r}}_{j})\hat{n}_{j}$ and $\hat{D}_{10}$=$\sum_{j=1}^{K} {u_{1}}^{*}({\bf{r}}_{j})u_{0}({\bf{r}}_{j})\hat{n}_{j}$. The steady state solution again gives the photon transmission in terms of the atomic variables as 
\beq
\hat{a}_{1}^{\dag} \hat{a}_{1} = \frac{\delta_{1}^{2} \hat{D}_{10}^{\dag} \hat{D}_{10} |\eta |^{2}} {([\Delta_{p} - (\hat{\omega} + \hat{\Omega})]^2 +  \kappa^2)([\Delta_{p} - (\hat{\omega} - \hat{\Omega})]^2 +  \kappa^2)} \label{a1}
\eeq
and reveals the nature of the many body quantum state of ultra cold atoms. 

Here, $\hat{\omega}$= $\frac{\delta_{1}}{2}(\hat{D}_{11}+\hat{D}_{00})$ 
and $\hat{\Omega} = \sqrt{\frac{\delta_{1}^{2} ({\hat{D}_{11} - \hat{D}_{00}})^{2}}{4} + \delta_{1}^{2} \hat{D}_{10}^{\dag} \hat{D}_{10}}$. 
From Eq.~(\ref{a1}) that the transmission spectra from the second cavity mode consists of two peaks, one at $\langle\hat{\omega}-\hat{\Omega}\rangle$ and the other at $\langle\hat{\omega}+\hat{\Omega}\rangle$  and thus, separated from each other by 2$\langle \hat{\Omega} \rangle$, which constitutes the normal mode splitting, which is also a measurable quantity. In the work of Mekhov et al. \cite{mekhov1, mekhov2}, the expectation value of the above defined photon number operators (Eq. (\ref{a0}) and Eq. (\ref{a1})) is taken in a Mott insulator state and superfluid state defined in the Fock space basis.
Mott insulator is a single Fock space state with well defined particle number at each site. In contrast to an insulating phase, in a superfluid phase, the atoms are delocalized and keep hopping from one site to another. Thus, a superfluid can be represented as a superposition of all possible Fock space basis vectors corresponding to all possible distributions (arrangements) in which a system of $N$ particles distribute among $M$ sites.
The corresponding  many body wave functions are given by, 
\bea | \Psi_{MI} \rangle & = & |n,n,n,....,n\rangle   \label{mott}  \\
| \Psi_{SF} \rangle & =  & \frac{1}{M^{N/2}} \sum_{i=1}^D \sqrt{\frac{N!}{n_{1}!n_{2}!...n_{M}!}} | n_{1}, n_{2}, .., n_{M}\rangle	\label{WSF}
\eea 
where the summation is over all possible arrangements and $D$ is the total number of such Fock space basis vectors, if we consider $N$ particles and $M$ sites, $D=\frac{(N+M-1)!}{N!(M-1)!}$.

\subsection{Extension to a general class of insulator to superfluid transition}
An insulating state of such ultra cold atoms in optical lattice is characterized by a well defined number of atoms in each site and is achieved when the relative magnitude of tunneling amplitude from one optical lattice to its neighboring sites goes to zero.
For purely 
onsite interaction, the only insulating state that can be formed is a Mott insulator with same number of 
atoms in each site. For long range interactions, whose 
effect can be depicted minimally by including a nearest neighbor interaction term ( Eq.~(\ref{ebhm})) or even higher order 
terms such as next nearest neighbor etc., density wave insulating phases with different number of atoms at lattice sites are possible. 
As the tunneling amplitude between the neighboring sites is enhanced by varying the strength of the optical lattice 
potential, any of the insulating phases will make a transition to the superfluid phase. However, depending on the starting insulating state, the nature of these transitions and intermediate phases can vary. This provides us with a motivation 
to propose a variational ansatz that describes a superposition of an insulating phase $|I\rangle$ and a superfluid $|SF\rangle$, given by
\begin{equation}
\left|\psi\right\rangle=\alpha\left|I\right\rangle+f(\alpha)\left|SF\right\rangle
\label{waf}
\end{equation}
where,  
$|I\rangle $ denotes an insulating phase whose Fock space structure is well defined. It must be mentioned that this variational 
ansatz is independent of the parameters that appeared in the parent Hamiltonian such as the one defined in Eq. (\ref{ebhm}) and therefore, does not define the true ground state of this system. However, this works as a good starting point from where we can  understand 
how cavity QED method can distinguish between various insulating states and their transition to superfluidity within an analytical framework. We checked separately that the 
that the results obtained using the above variational ansatz matches well with the ones obtained using exact diagonalization and perturbation theory approach \cite{notes}. 


The above state in Eq. (\ref{waf}) obeys the normalization condition {\it  i.e.} $\left\langle\psi\right.\left|\psi\right\rangle$=1; and the boundary condition that if $\alpha=1, f(\alpha)=0$, the system exists as 
 perfect insulating  phase; and when $\alpha=0, f(\alpha)=1$, the system is in a pure SF phase. Normalization condition gives  
\begin{equation}
f(\alpha)=- \alpha c_{I}+\sqrt{1+(c_{I}^{2}-1)\alpha^2},
\end{equation}
where $c_{I}$=$\left\langle SF\right.\left|I\right\rangle$.  
The cavity transmission for the state (Eq. (\ref{waf})) will be according to  Eq. (\ref{pn}) and can be calculated as,
\beq 
\langle \psi|\hat{a}^+_l\hat{a}_l |\psi\rangle= (\alpha^2+2f(\alpha)\alpha c_{I})\langle I |\hat{a}^+_l\hat{a}_l|I\rangle+(f(\alpha))^{2}\langle SF |\hat{a}^+_l\hat{a}_l|SF\rangle
\label{ansatz}
\end{equation}
We vary the value of $\alpha$ between $[0,1]$ and using the above equation, find out the cavity transmission spectra for a set of many body states that continuously interpolate between a  given insulating state and a superfluid state. 

An insulating state is essentially a crystalline state characterized by a certain lattice translational symmetry, therefore, we consider the cavity transmission spectrum under a lattice translation. We shall see that this reveals the nature of discrete lattice symmetry that the
system possesses. Recently, it has been shown \cite{Adhip} that by changing the angle between the cavity mode and the optical 
lattice, it is possible to couple the light with a subset of illuminated optical lattice sites in a selective manner. This allows one to explore the Fock space structure of the quantum many body states under certain condition. In this work, we show that 
a combination of translation along the optical lattice axis and the above mentioned angular displacement between the cavity 
mode and the optical lattice axis, can be used as a powerful technique to explore the distribution of atoms in different sites for various insulating states revealing their respective  discrete translational symmetry. 
Accordingly, in the next section, we shall first consider what will happen to the cavity transmission spectrum as (a) the cavity is 
shifted along the optical axis (translation), (b) the orientation of the cavity with respect to the lattice is changed and (c) finally what will happen when these two methods are combined.

\section{Results for a single cavity mode}
\label{scm}
\subsection{Comparison between different insulating phases}
\begin{figure}[h!]
\includegraphics[width=14cm,height=6cm]{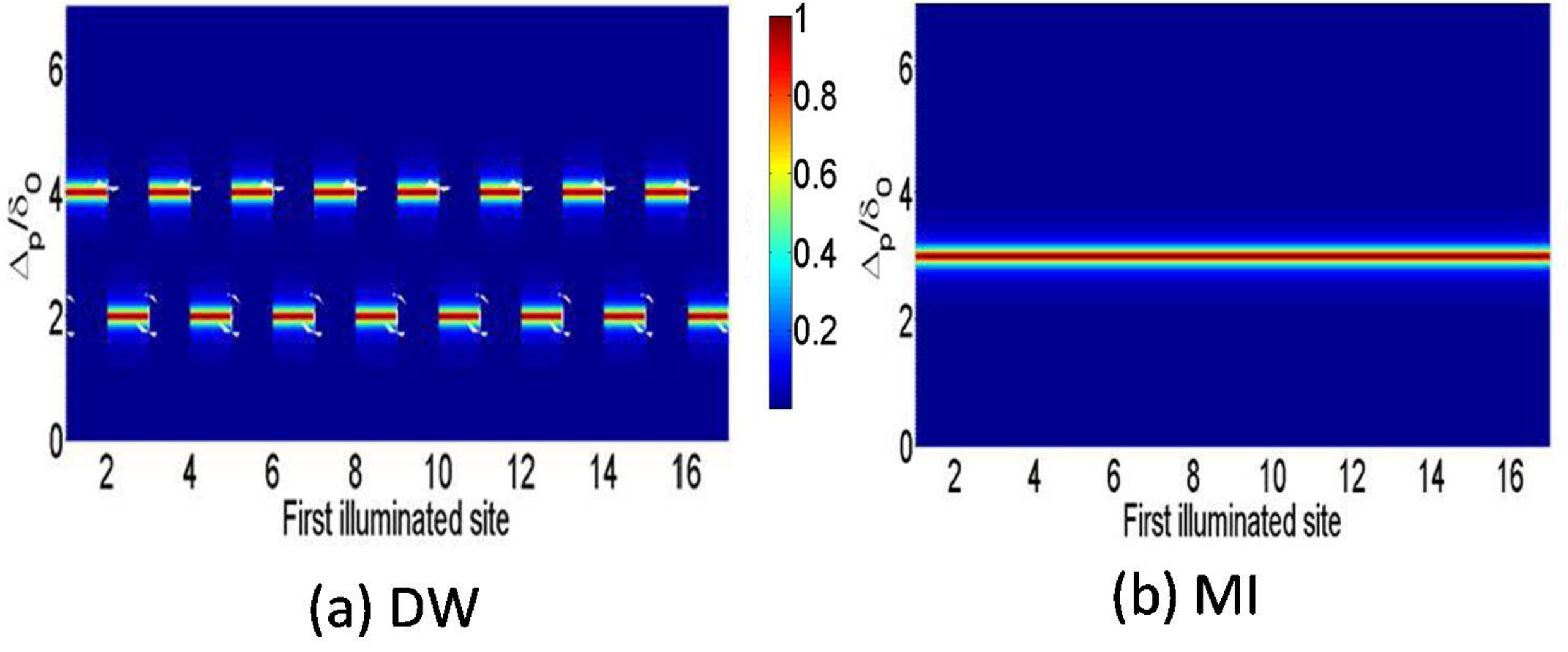}
\caption{(Color online) Transmission spectra as a function of the translational shift of the cavity for (a) DW and (b) MI state, for a system of $N$=20 particles, $M$=20 sites, $K$=3 illuminated sites. The color axis represents the transmitted number of photons given as $\langle \hat{a}_0^\dagger \hat{a}_0 \rangle/(|\eta|^2/\kappa^2)$}
\label{MIDWt}   
\end{figure}

A general example of an insulating state will be a two sublattice density wave (DW) state where two consecutive sites have different number of 
particles such that in the Fock space basis it can be written as  
\beq |I\rangle =|n_a,n_b,n_a,n_b,..\rangle . \label{DW} \eeq 
For such a state, 
\beq
c_{I}=\left\langle SF\right.\left|I\right\rangle=\sqrt{\frac{N!}{M^{N}(n_{a}!)^{(M/2)}(n_{b}!)^{(M/2)}}}
\eeq
Let us first start with the specific  DW state with $n_a= n_{0} + \Delta n$ and $n_{b}=n_{0} - \Delta n$ such that,  
\beq | I \rangle = |n_0+\Delta n, n_0-\Delta n,n_0 +\Delta n,n_0 -\Delta n,..\rangle \label{ins1} \eeq  
For $n_{0} = \Delta n=1$, this is a DW state $|2,0,2,0,2...\rangle$ and 
for $n_{0} =1$ and $\Delta n=0$, this corresponds to the Mott insulator state $|1,1,1,1..\rangle$. As a specific example, we consider a system consisting of $N=20$ particles, $M=20$ sites and illuminate a set of $K$=3 (odd)  sites using a single cavity mode. We are assuming standing wave cavity modes with a rectangular mode cross section, for which all the $K$ sites interact equally (for the case when $\theta_0=0$ and $\theta_0=90^\circ$) and other sites do not interact at all. We shift the cavity along the optical lattice through one lattice constant, keeping number $K$ always fixed and evaluate 
the transmission spectra at each step. For $\alpha$=1, the system can either be in a DW or MI state depending upon the competition between the onsite and nearest neighbor interactions. The result is plotted in Fig.~\ref{MIDWt}. 

\begin{figure}[htbp]
\includegraphics[width=14cm,height=5cm]{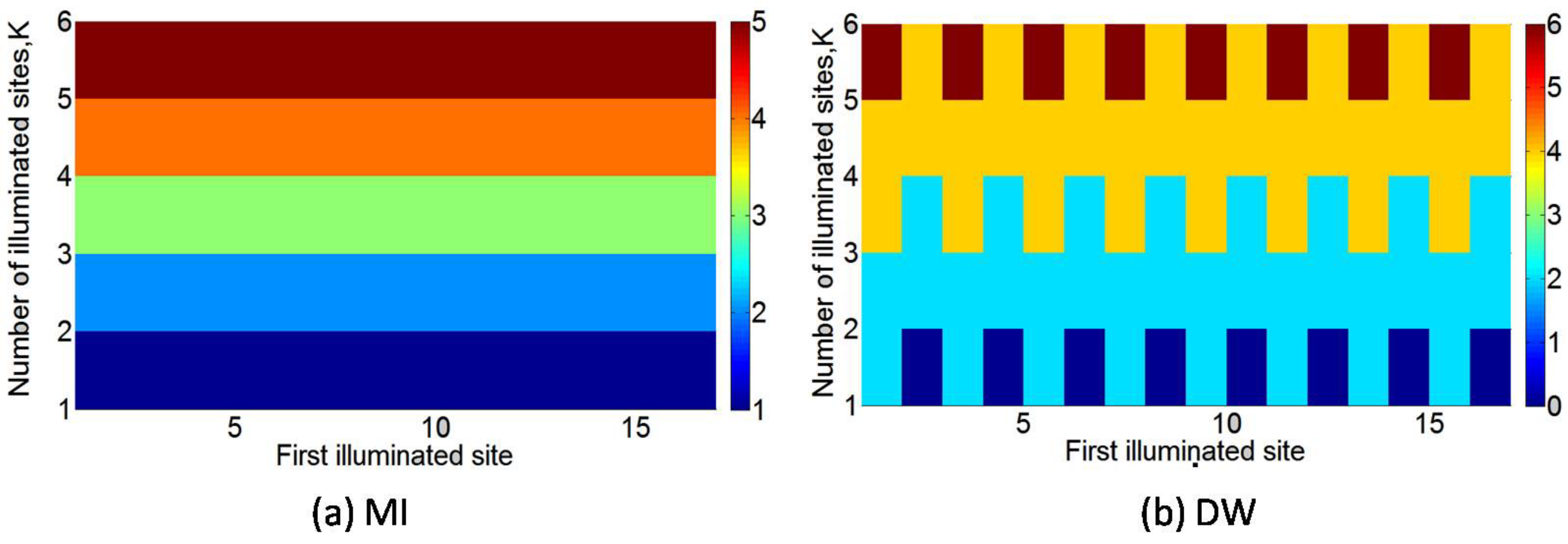}
\caption{(Color online) Variation of the transmission peak as a function of the translational shift of the cavity along the lattice axis (for $\theta_0$=0) and the number of illuminated sites (y axis) for (a) MI and (b) DW. The color axis shows the position of the transmission peak, $M=N=20$, in this case.}
\label{3DMIDW}   
\end{figure}

According to the figure, in DW phase ($|2,0,2,0..\rangle$), the position of transmission peak oscillates between 2 and 4 as one makes the translation. For an MI state, the position of the peak remains unchanged. Now referring to Eq. (\ref{a0}),
\bea
\hat{D}_{00}=n_0 K; \mbox{$K$=even} \nonumber\\
\hat{D}_{00}=n_0 K + (-1)^{i_{0}+1}\Delta n;  \mbox{$K$=odd}
\label{D_tr}
\eea
where $i_{0}$ is the site index of the first lattice site that is getting illuminated. As we translate the cavity step by step, $i_{0}$ increments by 1. Thus, the transmission spectra becomes,
\begin{equation}
\langle I|\hat{a_{0}}^+\hat{a_{0}}|I\rangle=\frac{|\eta|^2}{(\Delta_p-\delta_0 (n_0 K + (-1)^{i_{0}+1}\Delta n ))^2+\kappa ^2}
\label{a_0tr}
\end{equation}
For the  MI state, $\Delta n=0$  and the cavity transmission does not change with translation as the number of particles in the illuminated region stays the same. For the present case where $K$ is odd (=3), however,
for the DW state, as the cavity is translated, total number of illuminated atoms oscillates between 2 and 4 leading to the oscillation of the peak. The position of the peak signifies how many atoms are illuminated in the lattice, which for the above considered DW state, are either 4 or 2 at any translation step. 
For $K$=even number, transmission spectra of a MI and DW should be the same. This is because for even $K$, number of illuminated atoms for  MI and DW states considered is the same and hence the transmission peak is at the same position for both MI and DW and $\Delta n$ does not come into picture in the transmission spectra. A three dimensional plot of  such transmission spectrum of 
insulating DW and MI state for different values of $K$ is plotted in Fig. \ref{3DMIDW}. 
\subsection{Translation combined with angular shift}
\begin{figure}
\includegraphics[width=12cm,height=7cm]{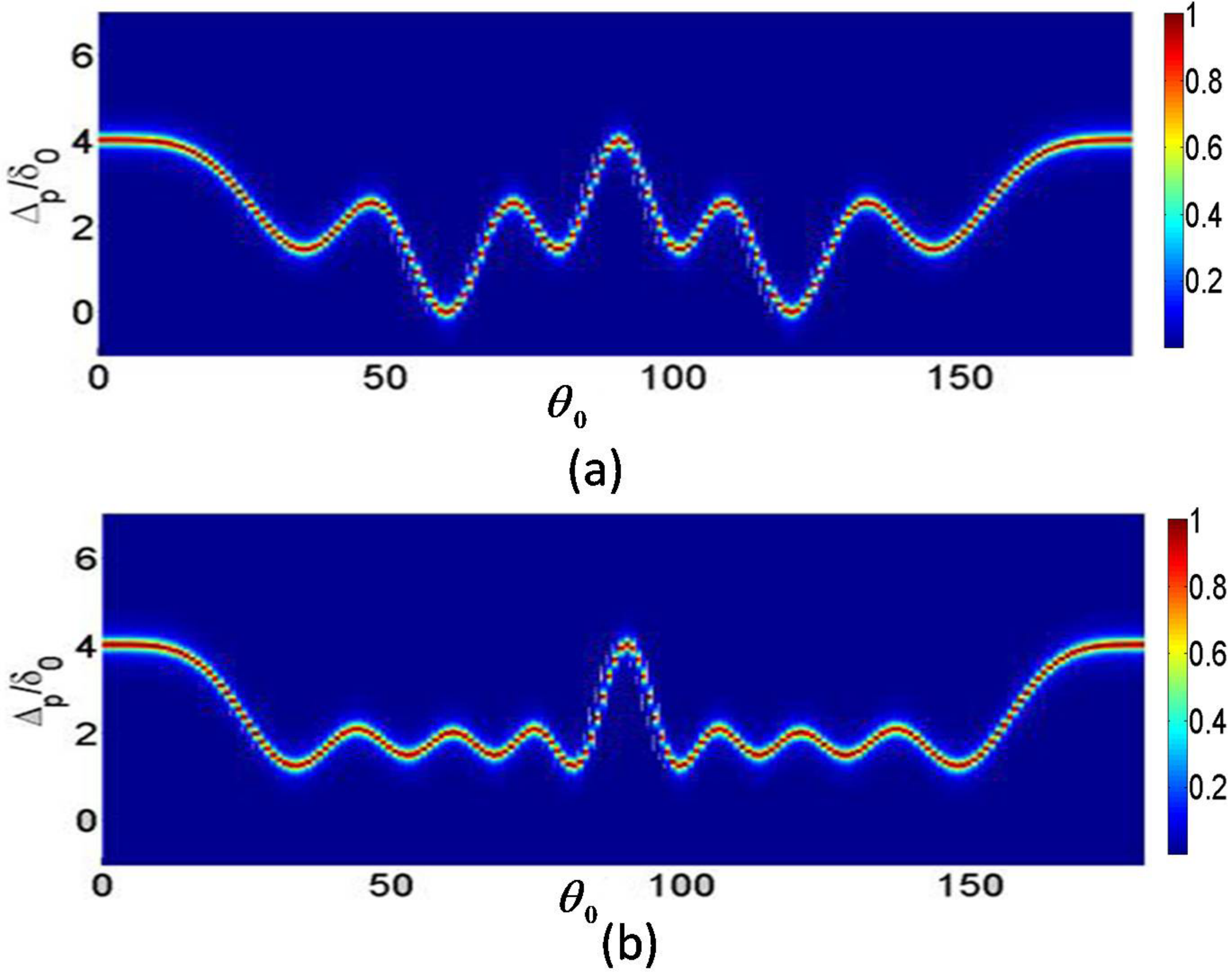}
\caption{(Color online) Variation in the transmission spectra as a function of the angle of the cavity with respect to the lattice axis ($\theta_0$) for (a) DW $|2,0,2,0..\rangle$ and (b) MI $|1,1,1,1..\rangle$ for $K$=4, $M=N=20$, color axis represents the transmitted photons given as ($\langle \hat{a}_0^\dagger \hat{a}_0 \rangle/(|\eta|^2/\kappa^2)$).}   
\label{mi_dw}
\end{figure}

The above method of identifying an insulating state from the change  
in the cavity transmission as a function of translation along the lattice axis however cannot distinguish  between 
an MI and DW  when the number of the illuminated sites are even. To make such distinction, the orientation of the cavity axis with the optical lattice direction ($\theta_0$) should be changed. 
The position of the transmission peak as we know is dependent upon the expectation value of $\hat{D}_{00} =\sum_{j=1}^K {u^* (\textbf{r}_j )u (\textbf{r}_j) \hat{n}_j } $. For standing waves, $u_0(\textbf{r}_j)=\mbox{cos}(j k_0 d + \pi)$, where $k_0=|\textbf{k}_{0} |\mbox{cos}\theta_0$ and $\phi$ is some initial phase shift (assumed to be zero here). The operator $\hat{D}_{00}$ becomes,
\begin{equation}
\hat{D}_{00} = \sum_{j=1}^K u_{0}(\textbf{r}_j)^{*} u_{0}(\textbf{r}_j) \hat{n_{j}} =  \sum_{j=1}^{K} \mbox{cos}^{2}(j\pi \mbox{cos}\theta_0) \hat{n}_{j} 
\end{equation} 
The eigenvalue of this operator for the state Eq. (\ref{ins1}) is given by
\bea
 E(\theta_0, K,n_0,\Delta n) & = & \frac{1}{2}n_0 \left[  K + \frac{\mbox{sin}(K\pi \mbox{cos}\theta_0) }{\mbox{sin}(\pi \mbox{cos}\theta_0)} \mbox{cos}((K +1)\pi \mbox{cos}\theta_0) \right] \nonumber \\ 
 & + & \frac{\Delta n}{4} \left[ 2+ (-1)^{K+1}\left[1+\frac{\mbox{cos}((2K+1)\pi \mbox{cos}\theta_0)}{\mbox{cos}(\pi \mbox{cos}\theta_0)}\right] \right]
 \label{Ecomp}
\eea
For an MI state, $\Delta n =0$ and for the DW state considered here, $\Delta n =1$. Therefore, the transmission spectrum given by 
\beq
\langle I|\hat{a_{0}}^+\hat{a_{0}}|I\rangle= \frac{|\eta|^2}{(\Delta_p-\delta_0 E(\theta_0, K,n_0,\Delta n))^2+\kappa ^2}
\eeq
will be different for the MI $|1,1,1,1..\rangle$ and DW $|2,0,2,0..\rangle$ state, irrespective of the number of illuminated sites $i.e.$ whether $K$ odd or even. This has been demonstrated in Fig.~\ref{mi_dw}.

Physically, this behavior can be understood and can be accounted to the fact that as we change the orientation of the cavity with respect to the lattice axis, this effectively changes the  wavelength or changes the $k$ vector of the cavity mode  \cite{Adhip}. Due to this, effectively, the total number of illuminated sites changes as the angle changes.

\begin{figure}[htbp]
\includegraphics[width=14cm,height=6cm]{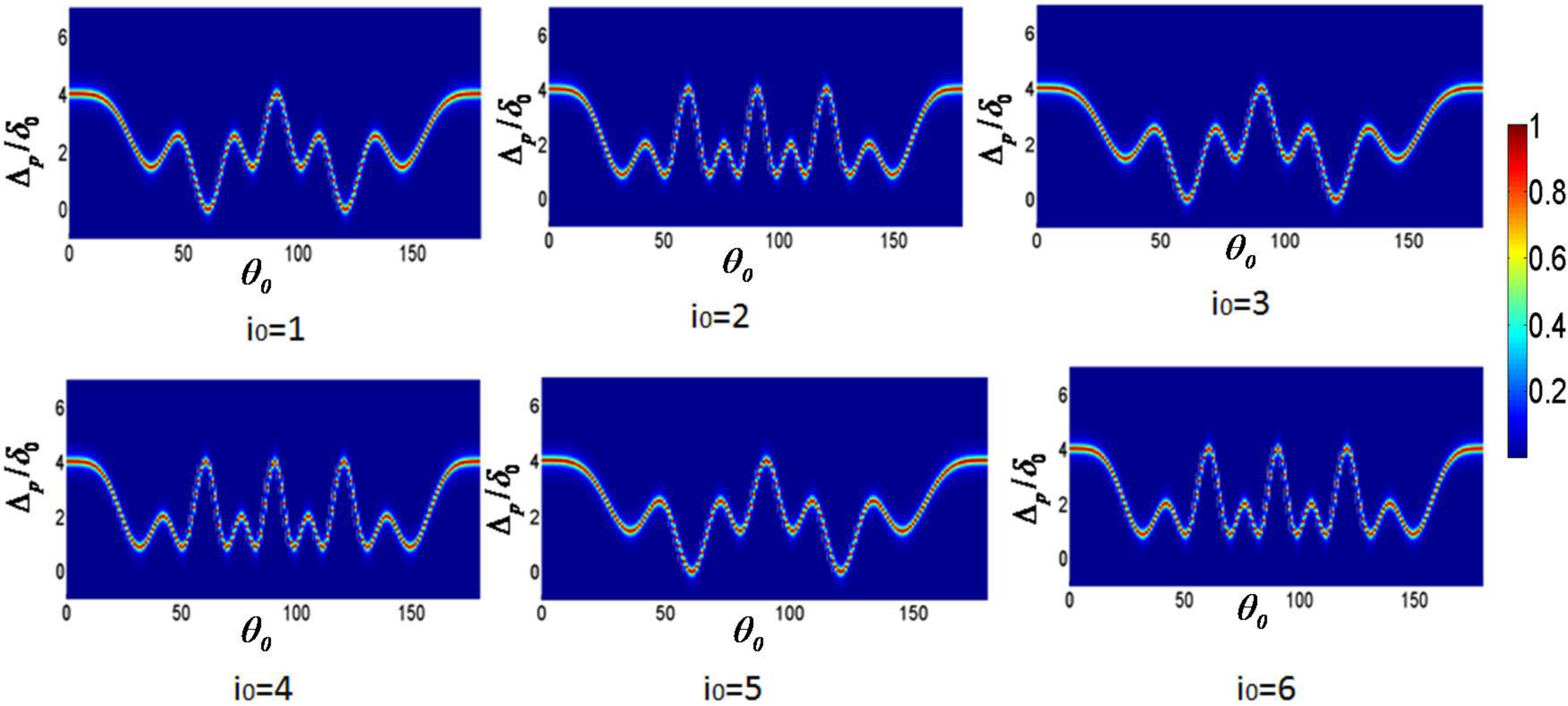}
\caption{(Color online) Variation in the transmission spectra as a function of the angle of the cavity with respect to the lattice axis ($\theta_0$) for a DW ($|2,0,2,0,..\rangle$) at different steps of translation, $N=M=20, K=4$.  The color axis represents $\langle \hat{a}_0^\dagger \hat{a}_0 \rangle/(|\eta|^2/\kappa^2)$.}   
\label{andw}
\end{figure}

It is therefore interesting to see that if we move one step further, namely change the orientation of the cavity to a particular angle with respect to the optical lattice and then translate the system along the optical lattice. 
The expression for the transmission spectra in Eq. (\ref{Ecomp}) would now get modified to
\bea
 E(\theta_0, K,n_0,\Delta n)& = & \frac{1}{2}n_0 \left[  K + \frac{\mbox{sin}(K\pi \mbox{cos}\theta_0) }{\mbox{sin}(\pi \mbox{cos}\theta_0)} \mbox{cos}((K +1)\pi \mbox{cos}\theta_0) \right] \nonumber \\ 
 & + & (-1)^{i_{0}+1}\frac{\Delta n}{4} \left[ 2+ (-1)^{K+1}\left[1+\frac{\mbox{cos}((2K+1)\pi \mbox{cos}\theta_0)}{\mbox{cos}(\pi \mbox{cos}\theta_0)}\right] \right]
\label{Ecompt}
 \eea
 where $i_0$ denotes the first illuminated site with respect to the lattice. The corresponding transmission spectra changes as a function of this translation shift and is shown in Fig. \ref{andw} for the DW phase. It can be seen from Eq. (\ref{Ecomp}) and Eq. (\ref{Ecompt}), that the first term in the expression for $ E(\theta_0, K,n_0,\Delta n)$ closely resembles that of a straight edge diffraction pattern and the second term which is proportional to $\Delta n$ signifies the contribution of the fluctuation in the mean particle density as a perturbation to this diffraction pattern. A more detailed discussion on the origin of this similarity appeared in an earlier work \cite{Adhip}. For the MI phase, the angular dependence of the transmission spectra is independent of the translation shift as expected (as $\Delta n$=0 for MI and hence $E(\theta_0, K,n_0,\Delta n)$ in Eq. (\ref{Ecompt}) becomes independent of the translational shift).  

For a DW, $\Delta n \neq 0$, and hence the contribution of the second term gives rise to additional features in the transmission spectra for a DW in the form of dips at certain angles. Physically, this issue can be thought of as diffraction of the photons of the standing wave cavity mode by the atoms present at the optical lattice which acts like a grating. In case of a DW state, all the lattice sites are not occupied uniformly and the deviation from the mean particle density gives rise to distinct features in the transmission spectra.

\subsection{Characterizing the MI to SF transition and DW to SF transition through single mode transmission}
\begin{figure}[htbp]
\includegraphics[width=14cm,height=7cm]{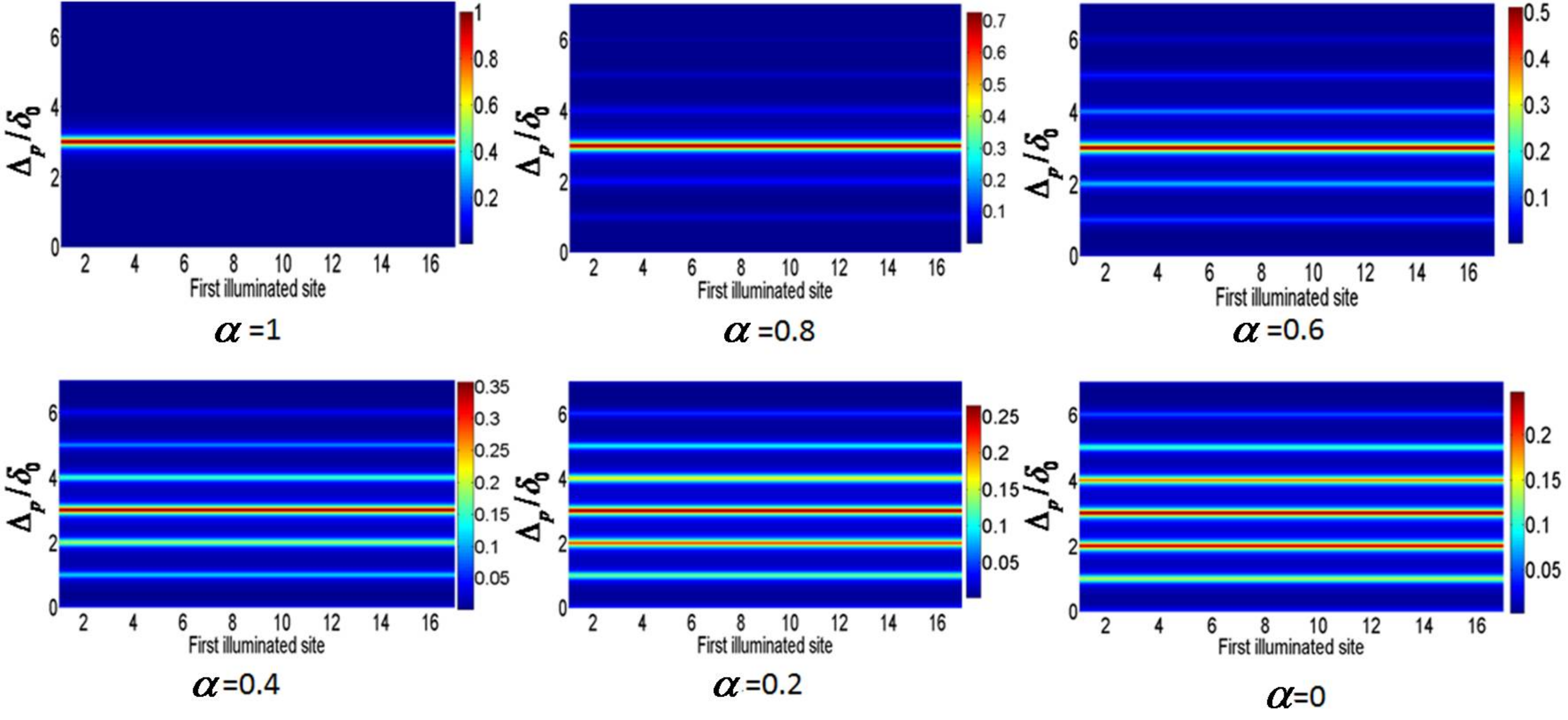}
\caption{(Color online) Variation in the transmission spectra as a function of the translational shift of the cavity for different values of $\alpha$ for a $|1,1,1,1..\rangle$ kind of Mott insulator state to SF phase transition. The color axis represents $\langle \hat{a}_0^\dagger \hat{a}_0 \rangle/(|\eta|^2/\kappa^2)$.}   
\label{1111tr}
\end{figure}

Following the standard variational technique, it is possible to obtain an acceptable value of the  parameter $\alpha$ by minimizing the energy given by Eq. (\ref{ebhm}) 
in the state Eq. (\ref{ansatz}) with respect to the variational parameter $\alpha$. However, the purpose of the introduction 
of the simple looking ansatz in (Eq. (\ref{ansatz})) is to see the gradual change of the transmission spectrum as $\alpha$ is being continuously varied rather than scouting for the true ground state of the system. We therefore content ourselves by looking at the transmission spectrum for the state Eq. (\ref{ansatz}) for various values of $\alpha$. 
The resultant transmission spectra for the DW to SF phase transition is shown in the plots of Fig.~\ref{2020tr} below.
\begin{figure}[h!]
\includegraphics[width=14cm,height=7cm]{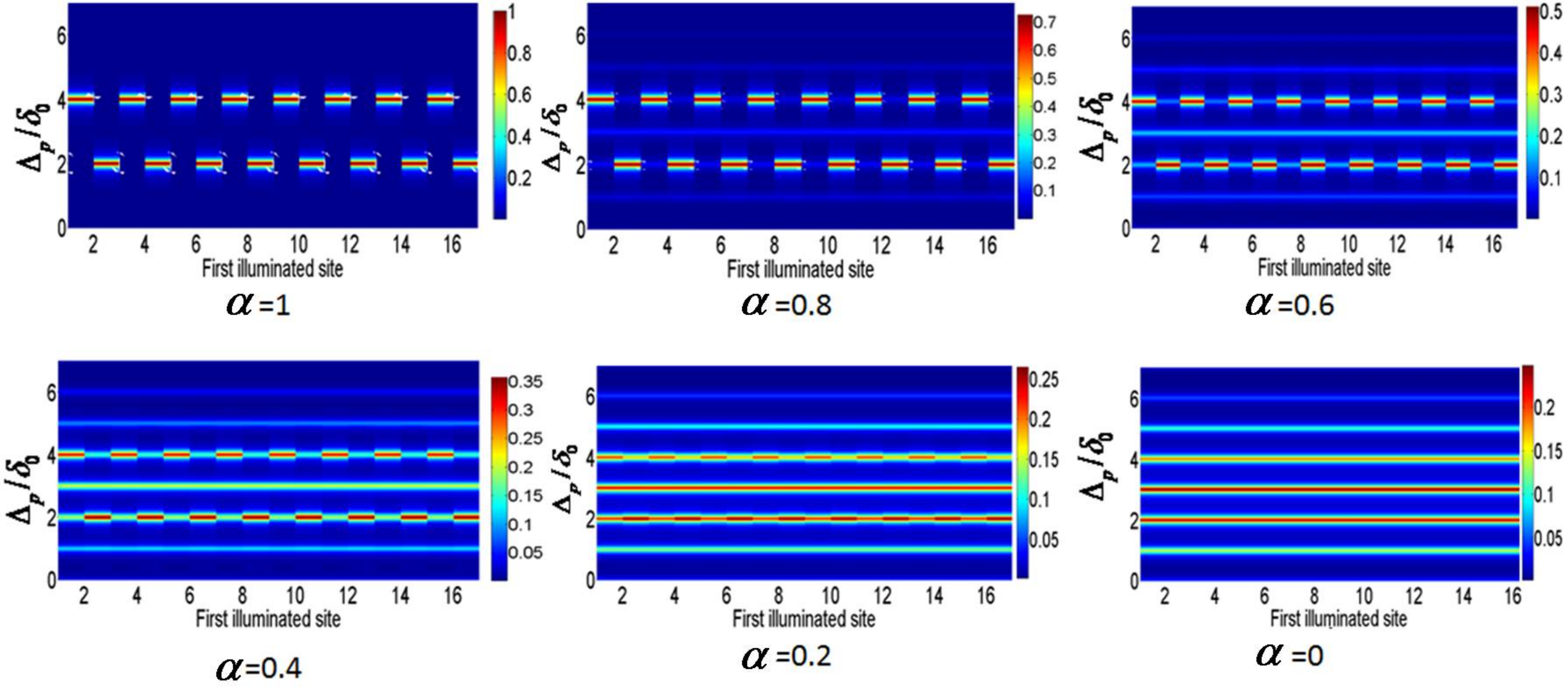}
\caption{(Color online) Variation in the transmission spectra as a function of the translational shift of the cavity for different values of $\alpha$ for a $|2,0,2,0,..\rangle$ DW state. Color axis represents $\langle \hat{a}_0^\dagger \hat{a}_0 \rangle/(|\eta|^2/\kappa^2)$. }
\label{2020tr}   
\end{figure}
As the value of $\alpha$ is reduced, the system is no longer in a  DW state, but a superposition of DW and SF state.
We already have the expression for transmission through a DW state. 
The transmission spectra of SF state  is given as 
\beq
\langle SF|\hat{a_{0}}^+\hat{a_{0}}|SF\rangle=\frac{1}{M^{N}} \sum_{i=1}^D \frac{N!}{n_{1}!n_{2}!...n_{M}!} \langle n_{1}, n_{2}, .., n_{M}|\frac{|\eta|^2}{(\Delta_p-\delta_0 \Sigma_{j=1}^K \hat{D}_{00})^2+\kappa ^2}|n_{1}, n_{2}, .., n_{M}\rangle
\eeq
 where the index $i$ runs over all possible Fock space basis vectors constituting the superfluid. To summarize, as we reduce the value of $\alpha$ (i.e. $\alpha$ $<$ 1), other Fock states also come into picture and interfere in the transmission spectra through the SF part, resulting in the multiple peaks. At larger values of $\alpha$ the amplitude of these 'other' peaks is lesser than the peaks at 2 and 4  on the $y$-axis. As $\alpha$ approaches 0, we get a SF state, which is marked by the emergence of a larger amplitude central peak, and lowering the side peaks at 2 and 4, as is evident from the Fig.~\ref{2020tr}.
Simultaneously, the oscillations in the peaks at 2 and 4 disappear and the peaks are uniform as we translate the cavity along the lattice as $\alpha=0$ approaches. This is another indication of  the transition to a superfluid 
phase from the cavity transmission spectrum. The corresponding situation as one goes from a MI to SF state is depicted 
in Fig.~\ref{1111tr}.

\begin{figure}[htbp]
\includegraphics[width=13cm,height=8cm]{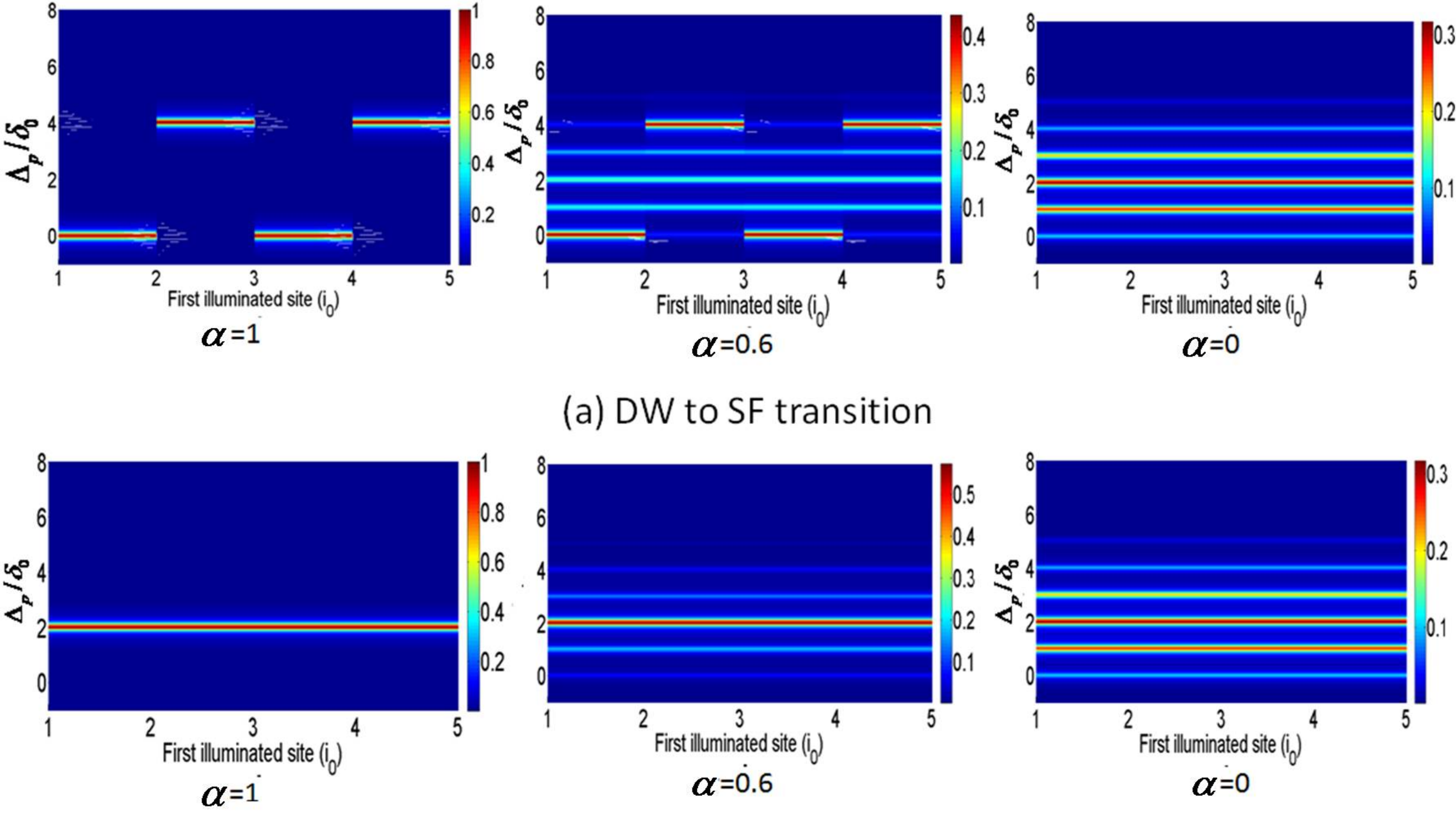}
\caption{(Color online) Translational shift at an angle of $60^\circ$, for (a) DW to SF phase transition and (b) MI to SF phase transition. In this case, $N=M=20$ and $K$=4 (even). The color axis represents $\langle \hat{a}_0^\dagger \hat{a}_0 \rangle/(|\eta|^2/\kappa^2)$.} 
\label{ants60}  
\end{figure}

In Fig. \ref{ants60}, we have again plotted such transmission spectrum as $\alpha$ is varied, but now making the angle between the cavity axis and the optical lattice $\theta^\circ=60^\circ$ and $K$=4 (even).

Here, even though the number of illuminated sites ($K$) is even, the position of the transmission peak oscillates between 0 and 4, which is a characteristic of a DW, and as we decrease the value of $\alpha$ to 0.6, there are multiple peaks, but the oscillation still persists; as $\alpha$ approaches 0, the prominent peak gets shifted at $\Delta_p/\delta_0$=2, which is signature of an SF phase. If 
a similar plot is made when a transmission is made from a  MI to SF phase, it will exhibit no such oscillations.

It is to be noted here that we are studying and analyzing the dispersion shift in the transmission spectra, which in turn, depends upon the number of atoms which are effectively getting illuminated by the cavity mode. Now if the number of atoms in the illuminated region fluctuates from time to time, the cavity mode gets a fluctuating frequency shift. This is the case for the pure superfluid state and also for an intermediate phase between the pure insulating and the superfluid state as represented by the superposed state (Eq. (\ref{waf})). Since, we are considering a subspace of $K$ sites, the measurement at any time, projects the state with fixed number of atoms. Any subsequent measurement, on a time scale shorter than the average tunneling time of atoms between the sites, will give the same result with the same value of dispersion shift. But if we repeat the experiment with a sufficient delay between the measurements to allow the atoms to tunnel and redistribute themselves in the $K$ sites, one should reproduce the spectra as shown in the plots~\cite{mekhov1}.

\section{Two mode analysis and  normal mode splitting}
\label{tcm}
\begin{figure}[htbp]
\begin{center}
\includegraphics[scale=0.4]{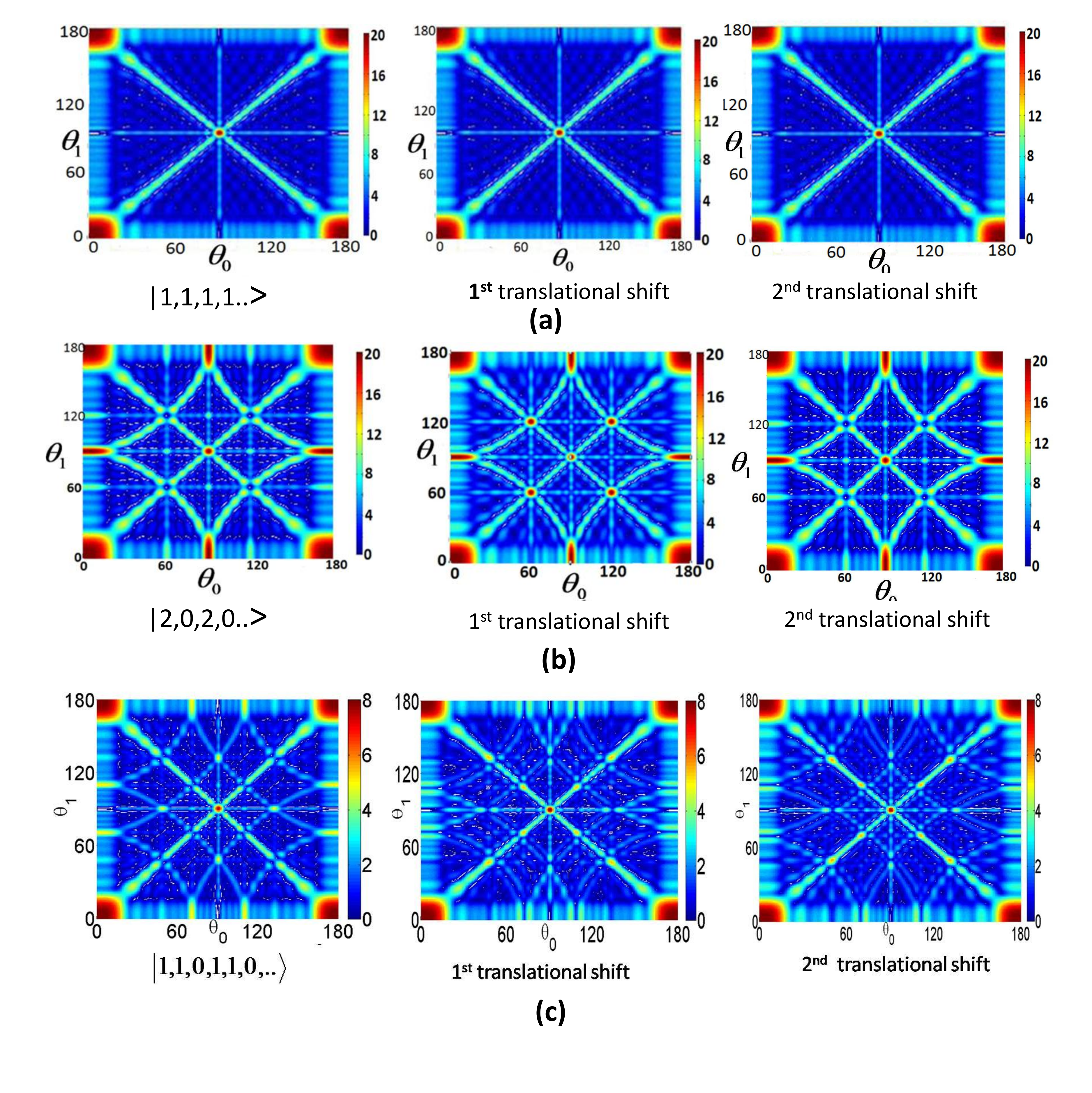}
\caption{(Color online) Variation of the normal mode splitting as a function of angles for different insulating phases (a) $|1,1,1,1..\rangle$ (MI) (b) $|2,0,2,0..\rangle$ (two sublattice DW) (c) $|1,1,0,1,1,0..\rangle$ (three sublattice DW). Different insulating phases show different NMS pattern, and under cavity translation show different behavior. The color axis gives the magnitude of the normal mode splitting.}
\label{nms_in}
\end{center}
\end{figure}

In the preceding discussion, we have focused on transmission spectrum for a single cavity mode. 
In this section, we will consider two cavity modes and consider the transmission from the second cavity through the expression given in Eq. (\ref{a1}). Such transmission demonstrates the normal mode splitting (NMS) because of the coupling between the two cavity modes. This 
NMS is a physically measurable quantity and has been used in various experiments~\cite{Bren1,nms1,nms2} to study  the scaling of atom-photon coupling strength $g$ and other aspects of atom-photon interaction. In the current work, we shall see how NMS can be used to differentiate between different insulating phases and their transition to the SF phase. 
%
For an insulating phase $|I\rangle$, which constitutes a single Fock space basis vector, the value of NMS can be found as
 \beq
\langle I|\hat{\Omega}|I\rangle=\frac{n_0}{2}\sqrt{\frac{(d_{00}-d_{11}+\Delta(d_{00}^{'}-d_{11}^{'}))^2}{4}+(d_{10}+\Delta d_{10}^{'})^2}
\label{spl}
\eeq
where
\bea
\Delta & = & \Delta n/n_0 \hspace{3cm}\nonumber\\
d_{00}& = & K+\frac{\mbox{sin}(K\pi \mbox{cos} \theta_0) \mbox{cos}((K+1)\pi \mbox{cos} \theta_0)}{\mbox{sin}(\pi \mbox{cos} \theta_0)}\nonumber \\
d_{00}^{'} & = & \frac{K}{2}+\frac{\mbox{sin}(K\pi \mbox{cos} \theta_0) \mbox{cos}((K+2)\pi \mbox{cos} \theta_0)}{\mbox{sin}(2 \pi \mbox{cos} \theta_0)}\nonumber \\
d_{11} & = & K+\frac{\mbox{sin}(K\pi \mbox{cos} \theta_1) \mbox{cos}((K+1)\pi \mbox{cos} \theta_1)}{\mbox{sin}(\pi \mbox{cos} \theta_1)}\nonumber \\
d_{11}^{'} & = & \frac{K}{2}+\frac{\mbox{sin}(K\pi \mbox{cos} \theta_1) \mbox{cos}((K+2)\pi \mbox{cos} \theta_1)}{\mbox{sin}(2 \pi \mbox{cos} \theta_1)}\nonumber \\
d_{10} & = & \frac{\mbox{sin}(K\pi \theta^{+})\mbox{cos}((K+1)\pi \theta^{+})}{\mbox{sin}(\pi \theta^{+})}+\frac{\mbox{sin}(K\pi \theta^{-}) \mbox{cos}((K+1)\pi \theta^{-})}{\mbox{sin}(\pi \theta^{-})}\nonumber \\
d_{10}^{'} & = & \frac{\mbox{sin}(S\pi \theta^{+})\mbox{cos}((S+2)\pi \theta^{+})}{\mbox{sin}(2\pi \theta^{+})}+\frac{\mbox{sin}(S\pi \theta^{-}) \mbox{cos}((S+2)\pi \theta^{-})}{\mbox{sin}(2\pi \theta^{-})}
\eea
with 
$\theta^{\pm}=\frac{\mbox{cos}\theta_0 \pm \mbox{cos}\theta_1}{2}$
and $S=K$ or $S=K-1$  respectively when $K$ is even or odd. As can be seen from the above equations that $d's$ are functions of $\theta_0$, $\theta_1$ and $K$ and hence would be same for any insulating phase. For the $|2,0,2,0..\rangle$ DW state, $\Delta n$=1, $n_0$=1, $\Delta$=1. For the DW state $|2,1,2,1,..\rangle$, $\Delta$=1/3, and for MI $|1,1,1,1,..\rangle$, $\Delta$=0. Since $\Delta$ is different for these insulating phases, they will show different NMS pattern. This is evident from Fig. \ref{nms_in}(a) and Fig. \ref{nms_in}(b), wherein we have plotted NMS as a function of the orientation of the cavities ($\theta_0$ and $\theta_1$) for the MI $|1,1,1,1..\rangle$ and $|2,0,2,0..\rangle$ DW phase, respectively. It can be seen from the plots that the NMS pattern for the MI phase and the DW phase is very different from each other at different orientations of the cavities. Since different insulating phases of cold atoms have different Fock space structure, they show different atom-photon interaction with the cavity modes at certain orientations of the cavities. This has a consequence that the two cavity modes will be coupled differently by different insulating phases and this is noticeable at several angles in the NMS measurements as shown in Fig. \ref{nms_in}.

 Now,  $|2,0,2,0..\rangle$ is a two sublattice DW phase, likewise $|1,1,0,1,1,0..\rangle$ is a three sublattice DW phase whose NMS pattern is plotted in Fig. \ref{nms_in}(c). It can be observed that the three sublattice DW states which arise due to other kinds of interactions between the ultracold atoms, have different kind of translational symmetry in comparison to the two sublattice DW states. In order to explore this, as a next step, we shift the cavities along the lattice axis with step size equal to one lattice constant and plot the NMS again as a function of angles when $K=6$ sites are illuminated. The result for this translational shift for different insulating phases is as shown in Fig.~\ref{nms_in}.

It is evident from this figure, that for a DW phase, the NMS pattern changes as we make a translation but for the MI, the behavior does not change even after translation. There are several geometries where this difference is very clearly demonstrated. Further, for the two sublattice DW phase ($|2,0,2,0..\rangle$),the NMS pattern repeats itself at the second translation shift (shown in Fig. \ref{nms_in}(b)) but for the three sublattice DW phase ($|1,1,0,1,1,0..\rangle$), the NMS pattern starts repeating from the third translational shift.   

\begin{figure}[htbp]
\begin{center}
\includegraphics[scale=0.6]{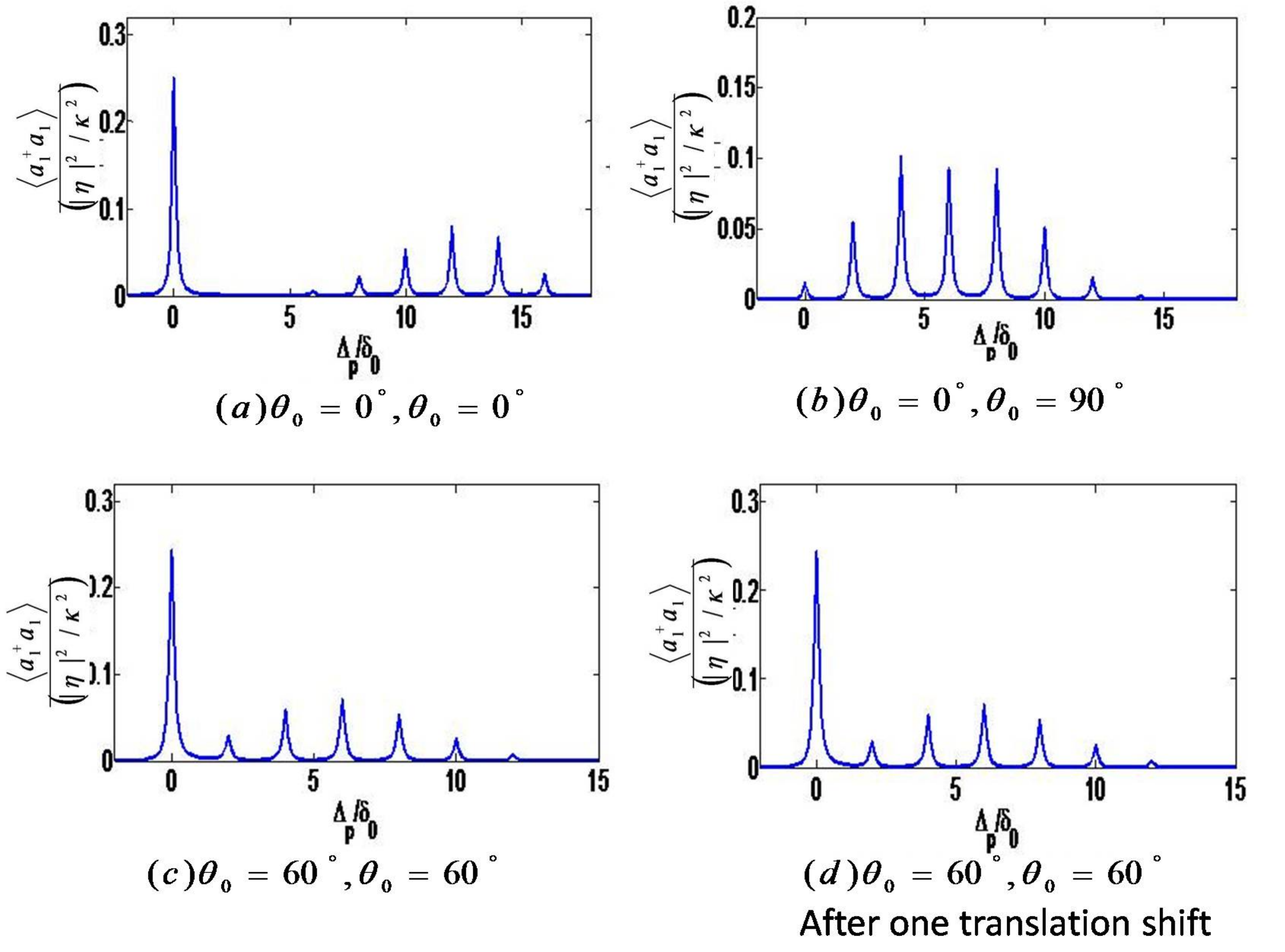}
\caption{(Color online) Transmission spectra for the two mode case for (a) $\theta_0=\theta_1=0^\circ$ (maxima) (b) $\theta_0=0^\circ,\theta_1=90^\circ$; (minima) (c) $\theta_0=\theta_1=60^\circ$; (d) $\theta_0=\theta_1=60^\circ$ (but after one translation) for the superfluid phase.}
\label{SF_nms}
\end{center}
\end{figure}
The corresponding normal mode splitting for a superfluid phase have less interesting features. This is because, a superfluid phase  which is a superposition of various Fock states,  
is a translationally invariant system.
 As an illustration, we consider the transmission spectra for the SF phase for different geometries as shown in Fig.~\ref{SF_nms}. The translational invariance is again seen in Fig.~\ref{SF_nms}(c),(d), 
like the case of single mode analysis.

In the present paper, we have studied far off resonant scattering of photons from the cold atoms in an optical lattice, assuming standing wave cavity modes and the atoms to be point like and have neglected back action of the scattered fields on the atoms. This is justified for the case of a deep lattice where the momentum transfer is insufficient to change the vibrational state of the atoms as long as not many photons are getting scattered. Further, in all of the above illustrations, we have translated the cavity along the lattice axis with a step size equal to one lattice spacing ($d$). Keeping in view the experimental difficulties that may arise during translating the cavity with this step size, we have also carried out simulations for translation shift equal to integer multiple of lattice spacing, which gives the desired result. The important aspect is that the cavity must be translated with a constant step size. It is also worth noting that for a two sublattice DW phase (such as $|2,0,2,2..\rangle$), the cavity must be translated with step size equal to any odd integral multiple of lattice spacing such that its lattice translational symmetry can be explored. It is possible that some of the insulating states may have some defects along the lattice axis, under which situation, the transmitted signal will become less clear. The quantitative measurements on the imperfections in the signal could then give valuable information about such defects.

\section{Conclusion}
\label{conclu}
To summarize, in this work we have shown that how a general class of insulating phases can be determined by studying the cavity transmission 
spectrum which can exhibit the discrete symmetry  associated with such phases in a very directly way. We also provide how this spectrum changes as one goes 
from such insulating phases to a superfluid phase which can be used to identify more exotic phases like supersolid.  
Of course, as a demonstration  we show the transmission spectrum with the help of a toy variational ansatz whereas the actual ground state of the real system will be more complicated.
However, once that ground state in the Fock space basis is determined, our method can be applied to it. If some of the states have some defects or the number of particles is not perfectly controlled, the signal will become less clear. Testing the in quantitative effect of such imperfections on the signal could give valuable information for experiments.
The implementation of our method depends on how experimentally 
one can translate or rotate the cavity with respect to the ultra cold atoms which are loaded inside the cavity as well as the relative size of the atomic condensate and the part of it that 
is illuminated. If this is achieved, according to our analysis,  the single mode transmission spectrum and the 
 normal mode splitting measurements offer us rich information about the various DW phases and their transition to a SF phase.  

JL is supported by a UGC fellowship given by Govt. of India.


\begin{thebibliography}{99}
\bibitem{notes}  For the Mott Insulator to Superfluid transition the agreement between the exact diagonalization and this variational method is excellent.
However  zero tunneling parameter (J) if the ground state is a density wave like state, the simulations using exact diagonalization method gives  $|2,0,2,0,..\rangle$ and $|0,2,0,2,..\rangle$ as two different types of ground states one of which is need to be projected out. Also for low value of tunneling the low values of tunneling, a perturbation theory can be employed and we checked that
the agreement with the variational ansatz is good.
\end{thebibliography}
\end{document}